\documentclass[prb,showpacs,twocolumn,aps,superscriptaddress,a4paper]{revtex4-1}

\usepackage{color}
\usepackage{dcolumn}
\usepackage{amsmath}
\usepackage{amssymb}
\usepackage{mathtools}
\usepackage{graphicx}
\usepackage{latexsym}
\usepackage{amsfonts}
\usepackage{amssymb}
\usepackage{comment}
\usepackage{natbib}
\DeclareGraphicsExtensions{.pdf,.gif,.jpg}

\newcommand{\be}{\begin{equation}}  
\newcommand{\ee}{\end{equation}}
\newcommand{\beq}{\begin{eqnarray}}  
\newcommand{\eeq}{\end{eqnarray}}  
\newcommand{\ket}[1]{| #1 \rangle}
\newcommand{\bra}[1]{\langle #1 |}
\newcommand{\ev}[2]{\langle #1 | #2 \rangle}
\newcommand{\mel}[3]{\langle #1 | #2 | #3 \rangle}
\renewcommand\Im{\operatorname{Im}}

\def\eps{\epsilon}
\def\w{{\omega}}
\def\im{{\mathrm{i}}}
\def\ex{{\mathrm{e}}}
\def\ud{\mathrm{d}}
\def\hc{\text{h.c.}}
\def\partiall{\stackrel{\leftarrow}{\partial}}
\def\unitmat{\mbox{$1$}}
\def\bunitmat{\mbox{$\mathbf{1}$}}
\def\bH{\mbox{\boldmath $h$}}
\def\bG{\mbox{\boldmath $G$}}
\def\hypf{{}_2{F}_1}

\begin{document}


\title{Distinguishing Majorana Zero Modes from Impurity States through Time-Resolved Transport}

\author{Riku Tuovinen}
\email{riku.tuovinen@mpsd.mpg.de}
\affiliation{Max Planck Institute for the Structure and Dynamics of Matter, 22761 Hamburg, Germany}

\author{Enrico Perfetto}
\affiliation{CNR-ISM, Division of Ultrafast Processes in Materials (FLASHit), Area della ricerca di Roma 1, Monterotondo Scalo, Italy}
\affiliation{Dipartimento di Fisica, Universit\`a di Roma Tor Vergata, Via della Ricerca Scientifica, 00133 Rome, Italy}

\author{Robert van Leeuwen}
\affiliation{Department of Physics, Nanoscience Center, FIN 40014, University of Jyv{\"a}skyl{\"a}, Finland}

\author{Gianluca Stefanucci}
\affiliation{Dipartimento di Fisica, Universit\`a di Roma Tor Vergata, Via della Ricerca Scientifica, 00133 Rome, Italy}
\affiliation{INFN, Sezione di Roma Tor Vergata, Via della Ricerca Scientifica 1, 00133 Roma, Italy}

\author{Michael A. Sentef}
\affiliation{Max Planck Institute for the Structure and Dynamics of Matter, 22761 Hamburg, Germany}


\begin{abstract}
We study time-resolved charge transport in a superconducting nanowire using time-dependent Landauer-B{\"u}ttiker theory. We find that the steady-state Majorana zero-bias conductance peak emerges transiently accompanied by characteristic oscillations after a bias-voltage quench. These oscillations are absent for a trivial impurity state that otherwise shows a very similar steady-state signal as the Majorana zero mode. In addition, we find that Andreev bound states or quasi-Majorana states in the topologically trivial bulk phase can give rise to a zero-bias conductance peak, also retaining the transient properties of the Majorana zero mode. Our results imply that (1) time-resolved transport may be used as a probe to distinguish between the Majorana zero mode and interfacial impurity states; and (2) the quasi-Majorana states mimic the transient signatures of the topological Majorana zero modes, indicating of the possibility of utilizing also the quasi-Majorana states in topological quantum computing as anyons with non-Abelian braiding statistics.
\end{abstract}

\maketitle


\section{Introduction}

Topological quantum computing~\cite{nayak_non-abelian_2008} is an active field of research based on the key idea to reduce quantum decoherence issues by using topologically protected states~\cite{Hasan2010,Qi2011}. Majorana fermions are their own antiparticles~\cite{Majorana1937}, and their condensed-matter analogues, Majorana bound state or Majorana zero mode (MZM), retain this feature~\cite{Kitaev2001}. They are thus considered to be promising candidates for technological advances in topological quantum computing~\cite{Kim2018,Lutchyn2018} since their non-abelian statistics allow performing quantum computation protected from environmental perturbations~\cite{ivanov_non-abelian_2001}. Even though various experimental signatures of MZM have been reported~\cite{Mourik2012,Finck2013,nadj-perge_observation_2014,Albrecht2016,Suominen2017,Gul2018,Wang2018,Liu2018}, a clear and unambiguous detection and the consequent control of these states has proven difficult so far. For example, other types of bound states~\cite{ruby_end_2015} or interfacial impurity states (IS) also give rise to in-gap states that contribute to transport or scanning tunneling spectroscopy signals. Therefore probes that unambiguously distinguish between MZM and IS are highly desirable.

Time-resolved spectroscopies allow for studying the dynamics of various processes such as charge transport \cite{ochoa_pumpprobe_2015}. For instance, in a transport setup exhibiting the MZM, there is no guarantee of instantly relaxing to a steady-state configuration once the junction has been ``switched on'' by, e.g., applying an external perturbation. In contrast, the nonequilibrium problems are often much richer and more interesting than equilibrium properties~\cite{Kundu2013,Covito2018,Bondyopadhaya2018}. This is especially relevant when nowaday transport measurements are pushing the temporal resolution to sub-picosecond regime~\cite{Prechtel2012,Cocker2013,Hunter2015,Rashidi2016,Cocker2016,Marguerite2017,Jelic2017,McIver2018}, and these ultrafast processes can be observed in real time.

In this paper we propose time-resolved transport as a probe in order to reveal the difference between topological MZM and ordinary IS. We simulate the transient dynamics in a quantum wire coupled to metallic electrodes using the time-dependent Landauer--B{\"u}ttiker formalism~\cite{Perfetto2008,svlbook,Tuovinen2014,Ridley2015,Tuovinen2016,Ridley2017} extended to include superconducting states in a Nambu spinor representation. By comparing the time-dependent build-up of a steady-state current after a sudden quench of the bias voltage between (i) a topological state with MZM and (ii) a non-topological state with trivial impurity end states, we discover that the dynamics for (i) and (ii) look significantly different. For case (i) the time-resolved current shows pronounced oscillations that shift with the applied bias voltage and correspond to transitions between the biased electrodes and the MZM. By contrast, for case (ii) no such oscillations are observed. In addition, we study the transient response of quasi-Majorana states (QMS) in the topologically trivial phase~\cite{Kells2012,Liu2017,Vuik2018,Moore2018,Yavilberg2019}, and we find that the QMS mimic the signatures of the MZM both in the stationary and transient regimes. The resulting Fourier spectra of the time-resolved current can therefore be used to identify the MZM or QMS.

\begin{figure}[t]
\centering
\includegraphics[width=0.45\textwidth]{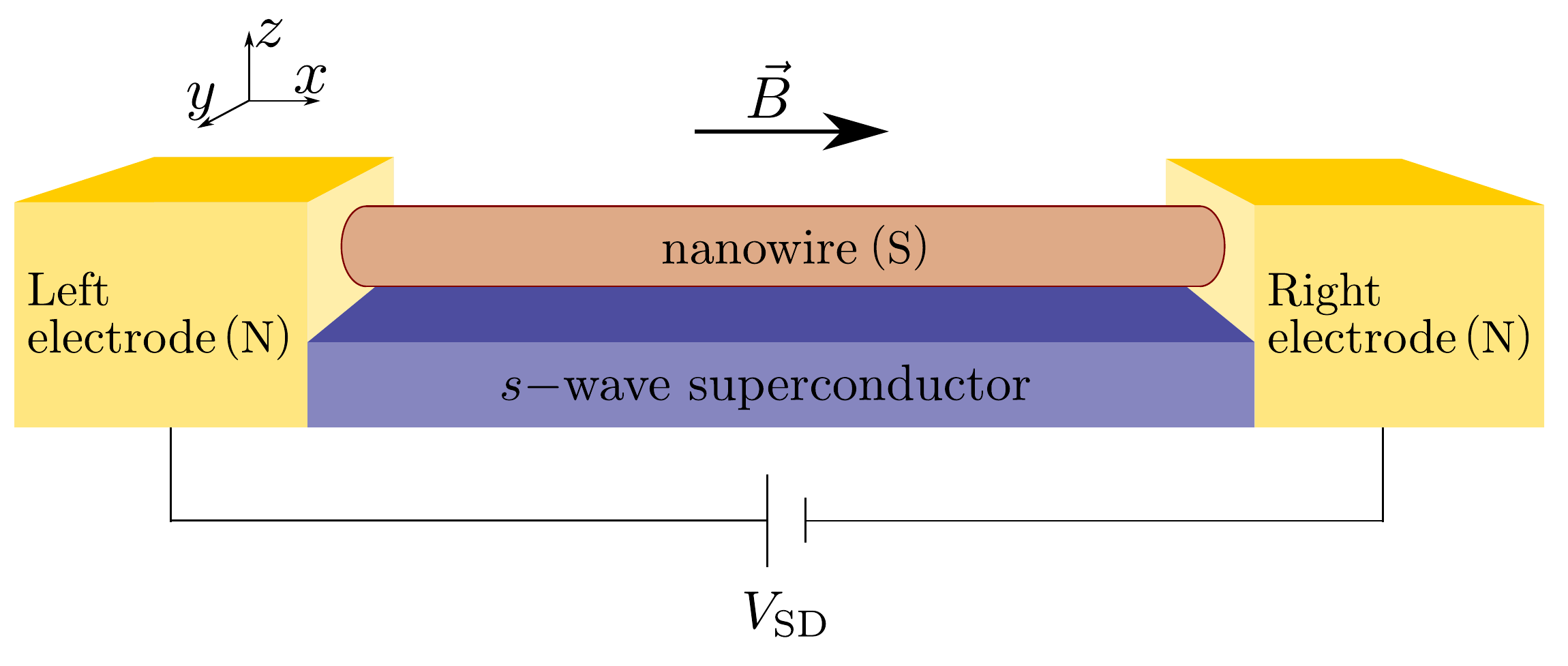}
\caption{A schematic NSN junction where two normal metal electrodes are connected to a nanowire where superconductivity is induced by the proximity effect from an adjacent $s$-wave SC. The electrodes are connected to a source--drain voltage $V_{\text{SD}}$. The magnetic field $\vec{B}$ orients the spins along the $z$ direction.}
\label{fig:setup}
\end{figure}

\section{Model and method}

We consider a normal metal--superconductor--normal metal (NSN) junction, see Fig.~\ref{fig:setup}. The superconducting central region of the junction is a nanowire in proximity to an $s$-wave bulk SC with order parameter $\varDelta$. The nanowire in addition features a strong spin-orbit interaction (e.g., InSb~\cite{Plissard2012,vanWeperen2015} or InAs~\cite{Fasth2007,Albrecht2016}) which favors aligning the spins along the $\pm y$ direction. An external magnetic field parallel to the nanowire breaks time-reversal symmetry and aligns the spins along the $\pm z$ direction, introducing a Zeeman splitting $V_{\text{Z}} = g\mu_{\text{B}} B/2$ where $g$ is the Land{\'e} factor and $\mu_{\text{B}}$ the Bohr magneton. A suitable combination of these effects has been shown to host a MZM in the nanowire, exponentially localized at the edges~\cite{Kitaev2001,Lutchyn2010,Oreg2010,Alicea2010,Stoudenmire2011,Mourik2012}. Specifically the infinite nanowire is in a topologically nontrivial phase for $V_{\text{Z}} > \varDelta > 0$  ~\cite{Lutchyn2010,Oreg2010,Stoudenmire2011}, from which MZM emerge in the case of a finite wire. For the present study, the specific structure of the electrodes, other than being a normal metal with relatively broad bandwidth (e.g., Au, Ag or Cu), is unimportant as we concentrate on the effects within the nanowire.

We write the total Hamiltonian as 
\be
\hat{H} = \hat{H}_e + \hat{H}_c + \hat{H}_w,
\ee
where the individual components for the electrodes and coupling are characterized by the single-particle energy dispersion in the electrodes $\epsilon_{k\lambda}$ and by the coupling matrix elements $T_{jk\lambda}$ between the states in the nanowire and the electrodes~\cite{Tuovinen2014}: 
\be
\hat{H}_e = \sum_{k\lambda}\epsilon_{k\lambda}\hat{c}_{k\lambda}^\dagger \hat{c}_{k\lambda}
\ee
and 
\be
\hat{H}_c = \sum_{jk\lambda}(T_{jk\lambda}\hat{c}_j^\dagger \hat{c}_{k\lambda} + \hc).
\ee
Here $k\lambda$ labels the $k$-th basis element in the $\lambda$-th electrode, and $j$ labels the atomic sites on the nanowire. The nanowire, in turn, is characterized by~\cite{Stoudenmire2011,Wu2012}
\beq\label{eq:wire}
\hat{H}_w & = & \sum_j {\Big[}-\frac{J}{2}{\large(}\hat{c}_j^\dagger \hat{c}_{j+1} + \hc{\large)} - (\mu-J)\hat{c}_j^\dagger \hat{c}_j \nonumber \\
& & -\frac{\alpha}{2}{\large(}\im \hat{c}_j^\dagger \sigma_2 \hat{c}_{j+1} + \hc{\large)} + V_{\text{Z}} \hat{c}_j^\dagger \sigma_3 \hat{c}_j \nonumber \\
& & +\varDelta {\large(}\hat{c}_{j\uparrow}\hat{c}_{j\downarrow} + \hc{\large)}{\Big]} ,
\eeq
where $J,\mu,\alpha,V_{\text{Z}}$, and $\varDelta$ are parameters for hopping, chemical potential, spin-orbit coupling, Zeeman splitting, and pairing potential, respectively. The operators $\hat{c}_{xs}^{(\dagger)}$ annihilate (create) electrons with spin $s\in\{\uparrow,\downarrow\}$ in a region specified by $x$. The spin indices are summed when suppressed and $\sigma_{2,3}$ are Pauli matrices. For indices $x,y$ belonging either to the electrodes or to the nanowire, the creation and annihilation operators satisfy the anticommutation relations $\{\hat{c}_{xs},\hat{c}_{ys'}^\dagger\}=\delta_{xy}\delta_{ss'}$.

At times $t>0$ the electrode energy levels are suddenly shifted, corresponding to a quench of the bias voltage, $\eps_{k\lambda}\to\eps_{k\lambda} + e V_\lambda$. For a two-terminal device ($\lambda\in\{\text{S},\text{D}\}$, see Fig.~\ref{fig:setup}) this out-of-equilibrium condition is defined by the source-drain voltage $V_{\text{SD}} = V_{\text{S}} - V_{\text{D}}$. The transport setup is considered \emph{partition-free}~\cite{Cini1980,Stefanucci2004,Ridley2018} meaning that the whole system is initially contacted in a global thermo-chemical equilibrium at unique chemical potential $\mu$ and at inverse temperature $\beta \equiv (k_{\text{B}}T)^{-1}$.

For a compact notation we introduce Nambu spinors~\cite{Nambu1960,Zeng2003,Xing2007} $\underline{\hat{\varPhi}}_x \equiv (\hat{\varPhi}_x^1 , \hat{\varPhi}_x^2 , \hat{\varPhi}_x^3 , \hat{\varPhi}_x^4 )^T \equiv (\hat{c}_{x\uparrow} , \hat{c}_{x\downarrow}^\dagger , \hat{c}_{x\downarrow} , \hat{c}_{x\uparrow}^\dagger)^T$, and the anticommutation relation is then understood componentwise $\{\hat{\varPhi}_x^\mu,(\hat{\varPhi}_y^\nu)^\dagger\} = \delta_{xy}\delta^{\mu\nu}$. Here we denote quantities in the Nambu$\otimes$spin space by an underline. This representation allows for writing the Hamiltonian for the nanowire in a Bogoliubov--de Gennes form~\cite{Bogoliubov1958,deGennes1966}
\be
\hat{H}_w = \frac{1}{2} \sum_j {\Big[} \underline{\hat{\varPhi}}_j^\dagger \underline{a}_j \underline{\hat{\varPhi}}_j + {\large(}\underline{\hat{\varPhi}}_j^\dagger \underline{b}_j \underline{\hat{\varPhi}}_{j+1} + \hc{\large)} {\Big]},
\ee
where we introduced on-site and nearest-neighbor contributions~\cite{Wu2012}
\begin{align}\label{eq:onsite}
& \underline{a}_j = \nonumber \\
& \begin{pmatrix} 
J-\mu + V_{\text{Z}} & -\varDelta & 0 & 0 \\
-\varDelta & \mu-J+V_{\text{Z}} & 0 & 0 \\
0 & 0 & J-\mu-V_{\text{Z}} & \varDelta \\
0 & 0 & \varDelta & \mu-J - V_{\text{Z}}
\end{pmatrix}_j \ , \
\end{align}
\be\label{eq:nn}
\underline{b}_j = \begin{pmatrix} 
-J/2 & 0 & -\alpha/2 & 0 \\
0 & J/2 & 0  & -\alpha/2 \\
\alpha/2 & 0 & -J/2 & 0 \\
0 & \alpha/2 & 0 & J/2 
\end{pmatrix}_j \ , \
\ee
respectively. The electrode and coupling parts of the Hamiltonian are then also expanded in the Nambu$\otimes$spin basis although they do not involve the SC pairing potential: 
\be
\hat{H}_e = \frac{1}{2}\sum_{k\lambda}\underline{\hat{\varPhi}}_{k\lambda}^\dagger \underline{\eps}_{k\lambda} \underline{\hat{\varPhi}}_{k\lambda}
\ee
with $\underline{\eps}_{k\lambda} = \epsilon_{k\lambda}\text{diag}(1,-1,1,-1)$ and 
\be
\hat{H}_c = \frac{1}{2}\sum_{jk\lambda}(\underline{\hat{\varPhi}}_j^\dagger \underline{T}_{jk\lambda}\underline{\hat{\varPhi}}_{k\lambda} + \hc)
\ee
with $\underline{T}_{jk\lambda} = T_{jk\lambda}\text{diag}(1,-1,1,-1)$.

By using the nonequilibrium Green's function approach~\cite{svlbook} we conveniently access both transient and steady-state responses in the setup above. The one-electron Green's function is defined as a contour-ordered tensor product of the spinor field operators~\cite{Zeng2003}
\be\label{eq:green}
\underline{G}_{xy}(z,z') = -\im \langle \mathcal{T}_\gamma[\underline{\hat{\varPhi}}_x (z)\otimes\underline{\hat{\varPhi}}_y^\dagger(z')] \rangle
\ee
where the contour-ordering operator $\mathcal{T}_\gamma$ is taken for the variables $z,z'$ on the Keldysh contour $\gamma$~\cite{svlbook}. The form in Eq.~\eqref{eq:green} automatically handles both normal and anomalous components of the Green's function~\cite{Stefanucci2010PNGF}. In the Supplemental Material~\cite{sm} we show that the equations of motion for the Green's function are exactly the same as those in Refs.~\cite{Tuovinen2014,Tuovinen2016}, and hence we derive in a similar fashion a closed expression for the time-dependent one-particle reduced density-matrix (TD1RDM) within the nanowire, $\underline{\rho}(t) \equiv -\im \underline{G}^<(t,t)$ from the lesser Green's function. In order to obtain a closed solution to the equation of motion we have described the electrodes within wide-band approximation, where the electronic levels of the nanowire are in a narrow range compared to the bandwidth of the electrodes. The coupling strength between the nanowire and the electrodes is characterized by the frequency-independent tunneling rate $\varGamma_\lambda$.

As the TD1RDM gives us full information on the local charge and current densities within the nanowire, we calculate the total current through the nanowire by considering a bond current between two atomic sites. In addition, the traditional bond-current operator has to be adapted to include the contribution from the spin-orbit coupling and from the SC pair potential~\cite{Solomon2010,Stefanucci2010,Winkler2017}. In the Supplemental Material~\cite{sm} we derive the following expression for the bond current between the sites $j$ and $j+1$ within the nanowire:
\beq\label{eq:bondcurrent}
I_{j,j+1} & = & 2 \Im {\Big[} -\frac{J}{2} {\large(} \langle \hat{c}_{j\uparrow}^\dagger \hat{c}_{(j+1)\uparrow} \rangle + \langle \hat{c}_{j\downarrow}^\dagger \hat{c}_{(j+1)\downarrow} \rangle {\large)} \nonumber \\
& & \hspace{26pt} - \ \frac{\alpha}{2} {\large(} \langle \hat{c}_{j\uparrow}^\dagger \hat{c}_{(j+1)\downarrow} \rangle - \langle \hat{c}_{j\downarrow}^\dagger \hat{c}_{(j+1)\uparrow} \rangle {\large)} \nonumber \\
& & \hspace{26pt} + \ 2\varDelta\sum_{m=1}^j \langle \hat{c}_{m\downarrow}\hat{c}_{m\uparrow} \rangle{\Big]},
\eeq
where $\langle \cdot \rangle$ denotes elements of the TD1RDM.

\section{Results}

\subsection{Emergence of the Majorana zero mode}

Using Eq.~\eqref{eq:bondcurrent} we calculate the (steady-state) current-voltage characteristics for nanowires of varying lengths. For the nanowire we choose the parameters $J=1$, $\alpha=0.5$, $V_{\text{Z}}=0.25$, $\varDelta=0.1$, and $\mu=0$~\cite{Wu2012}. This fixes the units to the hopping energy; if the values of this quantity are in the eV regime, then times are measured in the units of inverse hoppings which is on the order of femtoseconds. The coupling strength from the terminal sites of the nanowire to the electrodes is chosen such that the tunneling rate $\varGamma_\lambda = 0.01$. The bias voltage is applied symmetrically for the source and drain electrodes $V_S = -V_D \equiv V$, and we consider the zero-temperature limit.

\begin{figure}[t]
\centering
\includegraphics[width=0.5\textwidth]{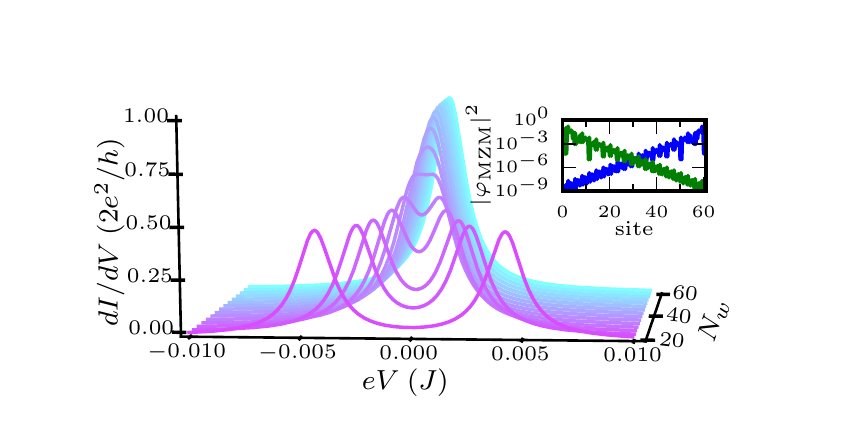}
\caption{Differential conductance versus applied bias voltage for nanowires of varying length $N_w$. The zero-bias peak builds up for sufficiently long nanowires ($N_w \gtrsim 50$). The probability density for the corresponding zero-energy modes shows exponential localization around the wire edges for $N_w=60$ (inset). Model parameters for the nanowire are $J=1$, $\alpha=0.5$, $V_{\text{Z}}=0.25$, $\varDelta=0.1$, and $\mu=0$.}
\label{fig:ivcurves}
\end{figure}

In Fig.~\ref{fig:ivcurves} we show the differential conductance against the applied bias voltage (around a low voltage window). We observe clearly how the MZM behaves as a ``half a fermion'' on both terminals of the nanowire leading to two peaks of half the conductance quantum. When the coupling between the MZM becomes weaker by elongating the nanowire, the two peaks merge into one at exactly zero bias voltage resulting in one conductance quantum. The inset of Fig.~\ref{fig:ivcurves} shows the exponential localization of the MZM for $N_w=60$. 

\subsection{Transient signature of the Majorana zero mode}

\begin{figure}[t!]
\centering
\includegraphics[width=0.5\textwidth]{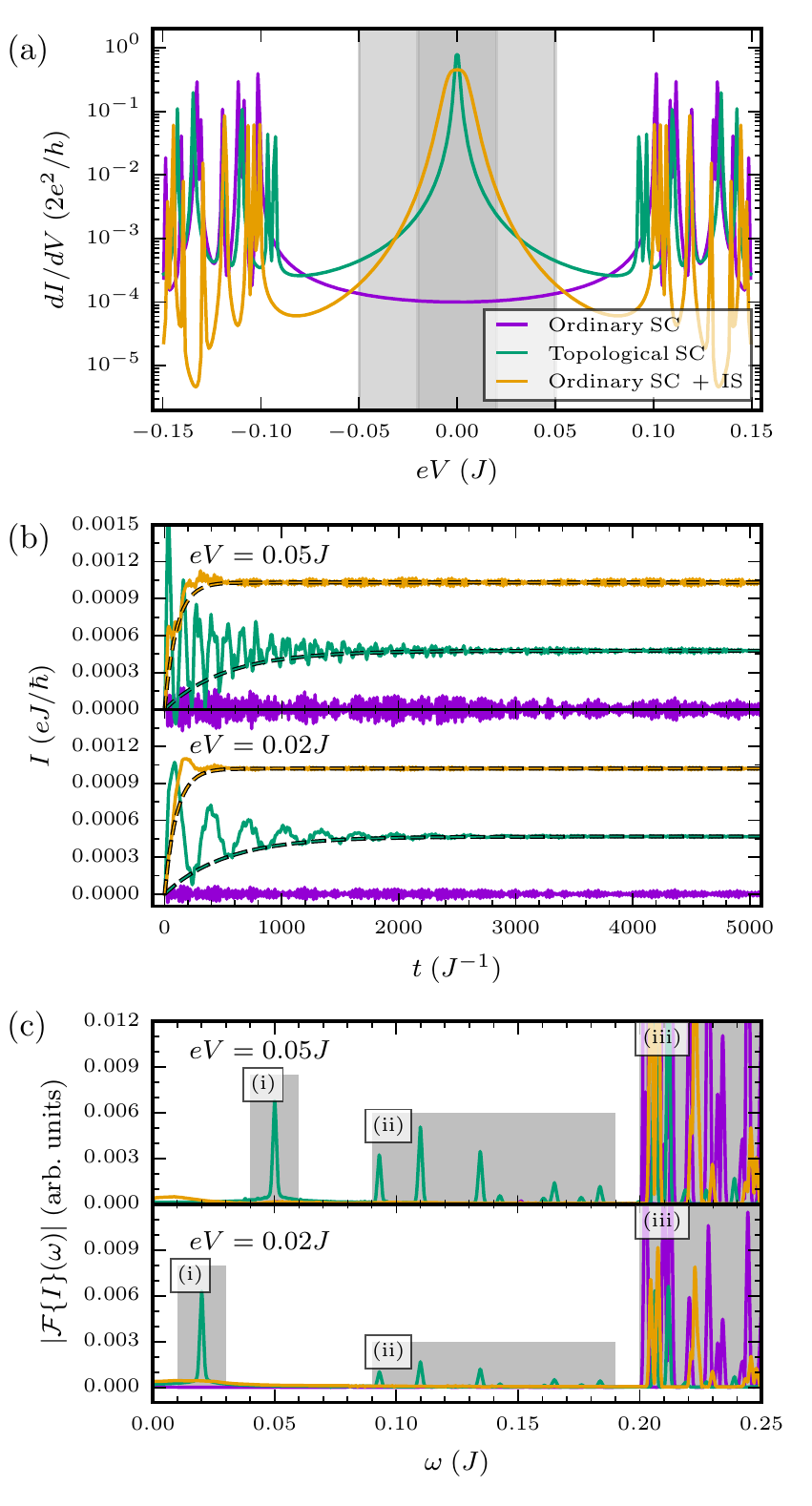}
\caption{(a) Differential conductance versus applied bias voltage for ordinary and topological SCs of length $N_w=50$. The shaded areas refer to the bias windows in panel (b). (b) Transient currents for applied bias voltages $eV=\{0.02,0.05\}J$. The dashed lines are given by $(1-e^{-\gamma t})I_{\mathrm{SS}}$ where $I_{\text{SS}}$ is the steady-state current and $\gamma$ is the decay rate, see text. (c) Fourier spectra corresponding to panel (b). The shaded areas (i)-(iii) result from different transitions, see text. Model parameters for the ordinary and topological superconductors are $J=1$, $\alpha=0.5$, $V_{\text{Z}}=\{0.0,0.25\}$, $\varDelta=0.1$, and $\mu=0$. For the ordinary superconductor with impurity states the model parameters are changed for the terminal sites of the nanowire as $\widetilde{\mu}=\widetilde{J} = 0.1J$, $\widetilde{\alpha} = 0.1\alpha$, and $\widetilde{\varDelta}=\widetilde{V}_{\text{Z}}=0$.}
\label{fig:transient}
\end{figure}

We evaluate transient currents through a $N_w=50$ nanowire by considering the two centermost sites in Eq.~\eqref{eq:bondcurrent}. We single out the MZM by applying a small bias window so that the oscillations in the time-resolved signal are only due to virtual transitions from the biased Fermi level of the electrode to the zero-energy mode in the nanowire. In Fig.~\ref{fig:transient}(a) we show the differential conductance for a $N_w=50$ nanowire for three different cases: (1) ordinary SC wire (same as Fig.~\ref{fig:ivcurves} but for $V_{\text{Z}}=0$), (2) topological SC wire corresponding to Fig.~\ref{fig:ivcurves}, and (3) an ordinary SC wire with an impurity state localized at its edges. We model the impurity states by modified tight-binding parameters~\cite{Robinson2008,Wehling2010} for the terminal sites in the nanowire, $j=\{1,N_w\}$ in Eqs.~\eqref{eq:onsite} and~\eqref{eq:nn}. More specifically, we use (modified parameters denoted by a tilde) $\widetilde{\mu}=\widetilde{J} = 0.1J$, $\widetilde{\alpha} = 0.1\alpha$, and $\widetilde{\varDelta}=\widetilde{V}_{\text{Z}}=0$. For our purposes the exact formulation is not too important as long as there is a separate state within the gap with different topological character compared to the MZM.

Importantly, while the steady-state $dI/dV$ signals of cases (2) and (3) look qualitatively similar (Fig.~\ref{fig:transient}(a)), the transient signals in Fig.~\ref{fig:transient}(b) for the three cases is qualitatively different. For the pristine wire without the MZM the current signal is zero on average due to there being no transport channels within the SC gap and the small bias window. When the IS is present as an in-gap state, the transient current rises rapidly but also saturates relatively fast to its stationary value within couple of hundred units of inverse hopping. (For hopping energies in the eV scale we have $J^{-1} \sim 0.658$~fs.) The IS is directly connected to the electrodes resulting in a strong hybridization and in a relatively fast decay of the transient. In contrast, the MZM at the edges of the nanowire have a different topological character being weakly coupled to each other although they are far apart, and even though the MZM is also directly connected to the electrodes, the hybridization of the MZM is weaker resulting in transient oscillations for thousands of time units, i.e., up to picoseconds. The decay rate can be approximated by the expectation value of the tunneling rate operator: $\gamma = \sum_{j=1}^2 \langle \varphi_{j} | \varGamma | \varphi_{j} \rangle$, where $\varGamma\equiv\sum_\lambda\varGamma_\lambda$ and $|\varphi\rangle$ are the IS or MZM eigenvectors, see the dashed lines in Fig.~\ref{fig:transient}(b). For identical wire-electrode coupling, the decay time $1/\gamma$ of the MZM transient current is roughly $5$ times the one of the IS.

Crucially, the MZM additonally shows transient current oscillations, unlike the IS. This striking difference between the MZM and IS cases is clearly seen by taking the Fourier transforms of the time-dependent signals, see Fig.~\ref{fig:transient}(c). The low-frequency regime shows pronounced peaks for the MZM case, and the frequency of the first peak exactly corresponds to the difference between the biased Fermi level of the electrode and the MZM (indicated by (i) in the figure). The analogous peaks in the case of the IS are strongly diminished. Importantly, we have also checked that by artificially increasing the decay time for the IS case by decreasing the wire-electrode coupling, there are still no pronounced transient oscillations in the IS case. Before entering the band of all possible transitions outside the SC gap ($\w\geq2\varDelta=0.2$, indicated by (iii) in the figure) we observe additional transitions between the MZM and states close to the gap edge (indicated by (ii) in the figure). These resonances remain independent of the applied voltage confirming that they result from intra-level transitions within the nanowire. Overall the transient features of the MZM are distinctly different from the IS.

\subsection{Comparison between Majorana zero modes and quasi-Majorana states}

Even though we found a clear distinction between trivial IS and the topological MZM, one may still wonder whether and how other in-gap states deeply in the topologically trivial regime for the same model of the nanowire would contribute to the time-resolved signal. Recently, it has been studied that in the parameter regime $\varDelta \ll V_{\text{Z}} < \mu$ the resulting in-gap QMS emerge without any additional impurities but by adding a smooth confining potential~\cite{Kells2012,Liu2017,Vuik2018,Moore2018}. These states can also be tuned arbitrarily close to zero energy, thereby mimicking the behaviour of the MZM.

We implement a confining potential within the nanowire as a simple function of the lattice coordinate $j\in[0,N_w)$ labeling the atomic sites on the nanowire:
\be
f_s(j) = 
\begin{cases}
\sin^2\left[\frac{\pi s j}{2 (N_w-1)}\right] , & j < \frac{N_w}{s}, \\
1 , & \frac{N_w}{s} \leq j < \frac{(s-1)(N_w-1)}{s}, \\
\cos^2\left[\frac{\pi s j}{2 (N_w-1)}\right] , & j \geq \frac{(s-1)(N_w-1)}{s}, \\
\end{cases}
\label{eq:conf}
\ee
where $s$ controls the smoothness at the edges. We then re-cast the values of the spin-orbit interaction and pair potential accordingly: $\alpha \to \alpha f_s(j)$ and $\varDelta \to \varDelta f_s(j)$. Large values of $s$ correspond to an abrupt hard-wall confinement where both $\alpha$ and $\varDelta$ remain constant (nonzero) throughout the nanowire (cf. previous subsections). For smaller values of $s$ the spin-orbit coupling and the induced superconductivity go to zero smoothly at the edges of the nanowire.

\begin{figure}[t!]
\centering
\includegraphics[width=0.5\textwidth]{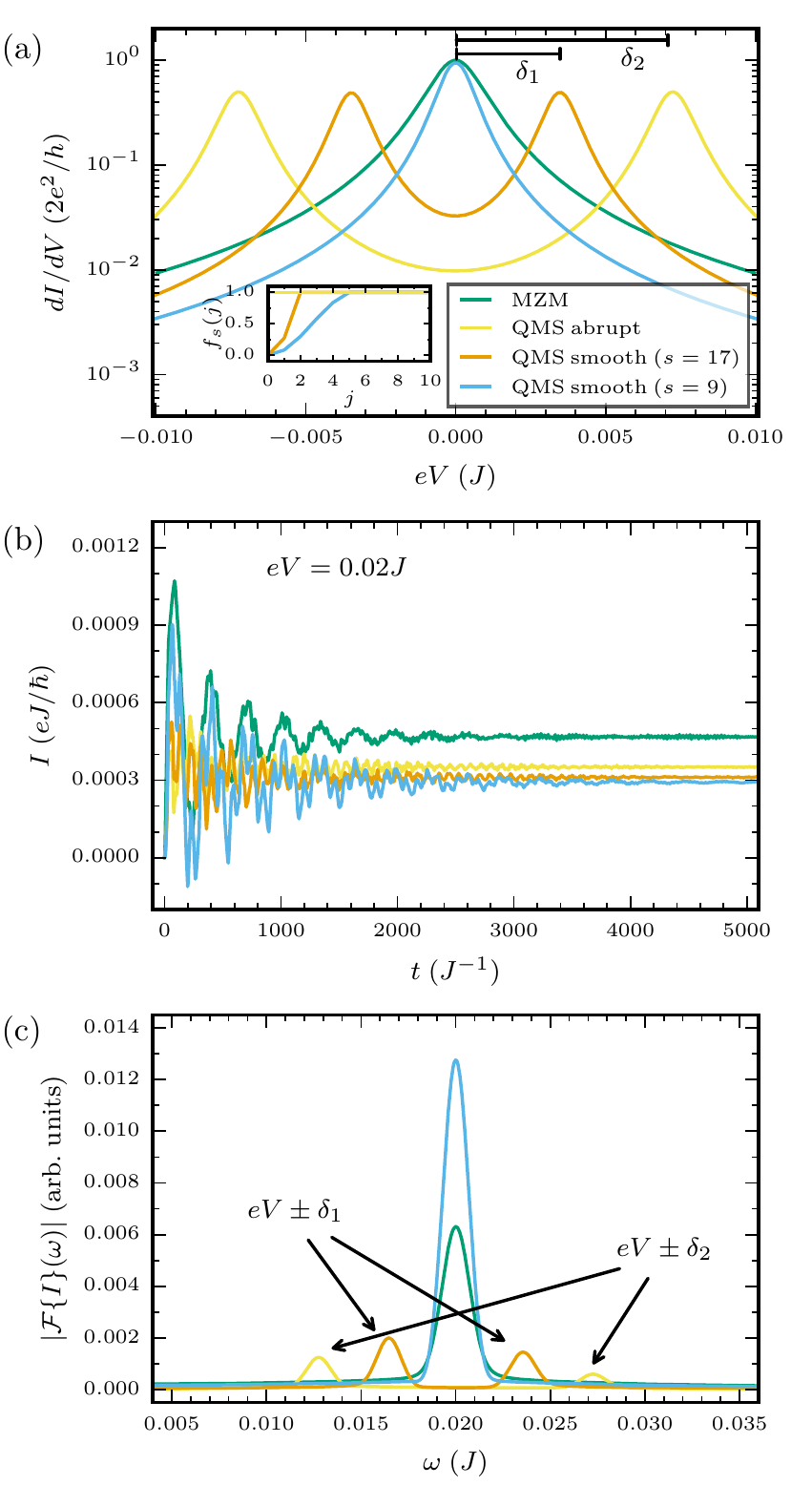}
\caption{(a) Low-voltage regime of the differential conductance for the MZM (cf. Fig.~\ref{fig:transient}) and QMS with varying confining potential. If the confining potential is not smooth enough, the QMS appear at nonzero energies $\pm\delta_{1,2}$. The inset shows the potential profile at the left end of the nanowire. (b) Transient currents for the separate cases in panel (a) when applying a bias voltage $eV=0.02J$. (c) Fourier spectra corresponding to panel (b). Model parameters for the QMS cases are $\alpha=0.5$, $V_Z = 1.2$, $\Delta=0.1$, and $\mu=2.0$.}
\label{fig:abs}
\end{figure}

To study the topologically trivial parameter regime we focus our discussion on three additional cases: For an $N_w = 50$ nanowire we set $\alpha=0.5$, $V_Z = 1.2$, $\Delta=0.1$, and $\mu=2.0$ both with abrupt ($s\to\infty$) and smooth ($s=\{17,9\}$) confinement potentials according to Eq.~\eqref{eq:conf}. In Fig.~\ref{fig:abs}(a) we show how the QMS is brought to a MZM-like state (peak at zero bias) by making the confining potential smoother. We have checked that other shapes for the potential profile do not modify the results qualitatively. The transient signature, see Figs.~\ref{fig:abs}(b) and~\ref{fig:abs}(c), of these states is also similar to the MZM: (1) The current oscillates with a dominant frequency corresponding to the lead-nanowire transition, and (2) the lifetime of the oscillations is similar or even longer compared to the MZM case. However, unless the QMS appears exactly at zero energy, the transient oscillations are suppressed, and the Fourier peak corresponding to the `smooth enough' case is considerably more pronounced.

\section{Conclusion}

We studied the time-dependent features of Majorana zero modes and quasi-Majorana states in a superconducting nanowire in contrast with trivial impurity bound states. The transient features related to MZM and QMS were found to be completely different than the ones resulting from a simple impurity model: The MZM and QMS transients were found to decay very slowly with a pronounced oscillation frequency due to a weaker hybridization of the MZM and QMS with the electrode states compared to the IS. This finding could be utilized in possible detection and identification of the MZM or QMS via ultrafast transport measurements \cite{Prechtel2012,Cocker2013,Hunter2015,Rashidi2016,Cocker2016,Marguerite2017,Jelic2017,McIver2018,sato_microscopic_2019}. 

We also found that even though the QMS are only protected by the smoothness of the confining potential (in contrast to the topological protection of the MZM), the QMS may still mimic the transient signature of the MZM. This effect could also be utilized by employing braiding schemes for the QMS in topological quantum computation~\cite{Vuik2018}. Since topological properties of the MZM should be robust against electronic interactions~\cite{Stoudenmire2011}, it would be a promising direction for future work to understand this effect for the QMS and how it might be manipulated and controlled.

In practice the sudden switch of the bias voltage employed here could be replaced by a short light pulse in the THz regime to excite the system away from its thermal equilibrium. In the case of an ultrashort laser excitation the current response of MZM or QMS could initially be suppressed and then recover transiently with the oscillations as a characteristic signature, similarly to the amplitude mode oscillations of laser-driven ordered phases~\cite{Kemper2015,Sentef2016,Kemper2017}. Together with ultrafast optical switching of chiral superconductors~\cite{dehghani_dynamical_2017,claassen_universal_2018} or nonequilibrium engineering of topologically nontrivial states of matter~\cite{oka_photovoltaic_2009,kitagawa_transport_2011,lindner_floquet_2011,jiang_majorana_2011,sentef_theory_2015,claassen_all-optical_2016,hubener_creating_2017,thakurathi_floquet_2017,topp_all-optical_2018} our findings highlight the great potential of ultrafast techniques for advances towards topological quantum computation.

\acknowledgments
R.T. and M.A.S. acknowledge funding by the DFG (Grant No. SE 2558/2-1) through the Emmy Noether program. R.v.L. acknowledges funding from the Academy of Finland (Project No. 317139). E.P. and G.S. acknowledge EC funding through the RISE Co-ExAN (Grant No. GA644076). E.P. also acknowledges funding from the European Union project MaX Materials design at the eXascale H2020-EINFRA-2015-1, Grant Agreement No. 676598 and Nanoscience Foundries and Fine Analysis-Europe H2020-INFRAIA-2014-2015, Grant Agreement No. 654360. G.S. also acknowledges Tor Vergata University for financial support through the Mission Sustainability Project 2DUTOPI.


%


\widetext
\newpage

\begin{center}
\textbf{\large Supplemental Material: Distinguishing Majorana Zero Modes from Impurity States through Time-Resolved Transport}
\end{center}
\setcounter{equation}{0}
\setcounter{figure}{0}
\setcounter{table}{0}
\setcounter{page}{1}
\setcounter{section}{0}
\makeatletter
\renewcommand{\theequation}{S\arabic{equation}}
\renewcommand{\thefigure}{S\arabic{figure}}
\renewcommand{\thepage}{S\arabic{page}}
\renewcommand{\bibnumfmt}[1]{[S#1]}
\renewcommand{\citenumfont}[1]{S#1}

\section{Transport setup and partitioning the Green's function}
Even though in the main text we considered a two-terminal device, the description readily allows for a more general treatment, and we now label by $\lambda$ an arbitrary number of electrodes. The central region $C$, for which we had the superconducting nanowire in the main text, can also take a more arbitrary shape. We only assume there to be no direct connection between any of the electrodes but the coupling is always through the central region. Then, the Hamiltonian for the full transport setup may be partitioned accordingly
\be\label{eq:hamiltonian-block}
\underline{\bH} = \begin{pmatrix}\underline{h}_{11} & 0 & \cdots & \underline{h}_{1C} \\
                                 0 & \underline{h}_{22} & \cdots & \underline{h}_{2C} \\ 
                                 \vdots & \vdots & \ddots & \vdots \\
                                 \underline{h}_{C1} & \underline{h}_{C2} & \cdots & \underline{h}_{CC}\end{pmatrix}
\ee
with $(\underline{h}_{\lambda \lambda'})_{kk'} = \underline{\epsilon}_{k\lambda}\delta_{\lambda \lambda'}\delta_{kk'}$ for the electrodes, and $(\underline{h}_{C\lambda})_{mk\lambda} = \underline{T}_{mk\lambda}$ for the couplings. For the central region, $\underline{h}_{CC}$, we may use the ``on-site'' and ``nearest-neighbor'' contributions [Eqs.~\eqref{eq:onsite} and~\eqref{eq:nn}], or consider some other arbitrary structure. We further denote the matrices for the full transport setup as boldface symbols. It is important to notice how the electrode blocks, $\underline{h}_{\lambda\lambda} = \underline{h}_{\lambda\lambda}(z)$, are different for the vertical and horizontal branches of the Keldysh contour due to the shift in energy levels at $t>0$. Also, we stress here that the block structure in Eq.~\eqref{eq:hamiltonian-block} does not refer to the Nambu$\otimes$spin space but it is of dimension $(N_e + 1) \times (N_e + 1)$ where $N_e$ is the number of electrodes. Each block then accounts for the individual dimension of the corresponding partition. The matrix elements in the Green's function in Eq.~\eqref{eq:green} (indices $x,y$ belonging either to the electrodes or to the central region) therefore label the transport setup in the same block form
\be
\underline{\bG} = \begin{pmatrix}\underline{G}_{11} & \underline{G}_{12} & \cdots & \underline{G}_{1C} \\ 
                                 \underline{G}_{21} & \underline{G}_{22} & \cdots & \underline{G}_{2C} \\\vdots & \vdots & \ddots & \vdots \\ \underline{G}_{C1} & \underline{G}_{C2} & \cdots & \underline{G}_{CC}\end{pmatrix}.
\ee

We may derive the equation of motion for the Green's function by
\be
\im\partial_z \underline{G}_{xy}(z,z') = \partial_z\left[\theta(z,z')\langle\underline{\hat{\varPhi}}_x(z)\otimes\underline{\hat{\varPhi}}_y^\dagger(z')\rangle  - \theta(z',z)\langle\underline{\hat{\varPhi}}_y(z')^\dagger\otimes\underline{\hat{\varPhi}}_x(z)\rangle \right] 
\ee
where the step function is defined on the Keldysh contour $\gamma$ according to the contour-ordering operator $\mathcal{T}_\gamma$~\cite{svlbook-s}. Evaluating the derivative gives
\be
\im\partial_z \underline{G}_{xy}(z,z') = \delta(z,z')\left\{\underline{\hat{\varPhi}}_x(z),\underline{\hat{\varPhi}}_y^\dagger(z')\right\} - \im\langle\mathcal{T}_\gamma[\im\partial_z\underline{\hat{\varPhi}}_x(z)]\otimes\underline{\hat{\varPhi}}_y^\dagger(z')\rangle\label{eq:eom-derive}
\ee
where the anticommutator gives simply $\delta_{xy}\underline{\unitmat}$ and the evolution of the spinor operator can further be derived from its equation of motion. Depending on which region the index $x$ belongs to (and the corresponding structure of the Hamiltonian in that region), the time-evolution of the field operator is completely specified. The equations of motion for the whole transport setup then take the matrix form~\cite{svlbook-s,Tuovinen2013-s,Tuovinen2014-s,Tuovinen2016-s}
\beq\label{eq:eom}
\left[\im \partial_z \underline{\bunitmat} - \underline{\bH}(z)\right]\underline{\bG}(z,z') & = & \delta(z,z')\underline{\bunitmat} , \\
\underline{\bG}(z,z')\left[-\im \partiall_{z'} \underline{\bunitmat} - \underline{\bH}(z')\right] & = & \delta(z,z')\underline{\bunitmat} ,
\eeq
which the Green's function satisfies being antiperiodic along the contour (Kubo--Martin--Schwinger boundary condition~\cite{Kubo1957,Martin1957}).

We see that the equations of motion are the same as those of Ref.~\cite{Tuovinen2013-s,Tuovinen2014-s}, hence we may in similar fashion, using the Langreth rules~\cite{LangrethRules,svlbook-s}, derive an equation for the equal-time lesser Green's function with indices on the central region $\underline{G}_{CC}^<$. This is a key quantity as it relates to the time-dependent one-particle reduced density-matrix (TD1RDM) by $\underline{\rho}_{CC}(t) = -\im \underline{G}_{CC}^<(t,t)$. From now on we will only discuss quantities in the subspace of the central region, so we will drop the subscript `$CC$'. The lesser Green's function at the equal-time limit is given by~\cite{Tuovinen2013-s}
\be\label{eq:final-glss}
\im\frac{\ud}{\ud t} \underline{G}^<(t,t)-[\underline{h}_{CC}(t),\underline{G}^<(t,t)] = -\left[\underline{G}^{\text{R}}\cdot\underline{\varSigma}^<+\underline{G}^<\cdot\underline{\varSigma}^{\text{A}}+\underline{G}^{\rceil}\star\underline{\varSigma}^{\lceil}\right](t,t)+\hc 
\ee
where the time-convolutions on the horizontal and vertical branches of the Keldysh contour are defined as $[f \cdot g](t,t) = \int_{0}^{\infty}\ud \bar{t} f(t,\bar{t})g(\bar{t},t)$ and $[f \star g](t,t) = -\im\int_{0}^{\beta}\ud \tau f(t,\tau)g(\tau,t)$. The superscripts R,A,$<,\rceil,\lceil$ refer to the retarded, advanced, lesser, right and left Keldysh components, respectively~\cite{svlbook-s,Tuovinen2013-s}. The embedding self-energy, $\varSigma$, accounts for the coupling between the central region and the electrodes~\cite{Tuovinen2014-s}.

We note that the left-hand side of Eq.~\eqref{eq:final-glss} corresponds to a Liouville-type of equation for the density matrix of an isolated central region whereas the right-hand side gives rise to an open transport setup as in connection to the electrode environment. The time-convolutions on the right-hand side can further be identified as source and drain terms, and the ones including the imaginary track of the Keldysh contour to include the initial contacting of the separate regions. Importantly, within the so-called wide-band approximation (WBA) for the embedding self-energy, Eq.~\eqref{eq:final-glss} becomes a closed equation for the equal-time lesser Green's function and the TD1RDM can be solved \emph{analytically}.

\section{Solution to the equation of motion}
In order to close the equation of motion we now describe the electrodes in the framework of wide-band approximation (WBA), where the electronic levels of the central region are in a narrow range compared to the electrode bandwidth. The validity of WBA has been discussed in, e.g., Refs.~\cite{Zhu2005,Maciejko2006,Verzijl2013,Covito2018}, and for the purpose of the present work (weak coupling of the central region to electrodes of large bandwidth), this is a well-justfied approximation. In frequency space the retarded Keldysh component of the embedding self-energy can then be written as
\be
\underline{\varSigma}_{\lambda,mn}^{\text{R}}(\w) = \sum_k \underline{T}_{mk\lambda}\frac{1}{\w-\underline{\epsilon}_{k\lambda}+\im\eta}\underline{T}_{k\lambda n} \approx -\im \underline{\varGamma}_{\lambda,mn}/2 .
\ee
The advanced component is given simply by conjugating this. The other components of the self-energy ($<,\lceil$) may further be derived from the retarded and advanced components~\cite{svlbook-s,Tuovinen2013-s}. The time-domain quantities in Eq.~\eqref{eq:final-glss} are then obtained by Fourier transforming. Looking at Eq.~\eqref{eq:final-glss} and the earlier work in Refs.~\cite{Tuovinen2013-s,Tuovinen2014-s} we may use the fact that the same equations have the same solutions, i.e., including the Nambu$\otimes$spin structure in the Hamiltonian of the central region (e.g., spin-orbit coupling, Zeeman splitting and pairing field) adds no extra complication to the evolution of the Green's function. The only difference is in the Nambu$\otimes$spin structure of the matrices. 

It is useful to introduce a nonhermitian effective Hamiltonian $\underline{h}_{\text{eff}} = \underline{h}_{CC} - \im\underline{\varGamma}/2$ for which the left and right eigenvalue equations are
\be
\bra{\varPsi^{\text{L}}}\underline{h}_{\text{eff}} = \eps\bra{\varPsi^{\text{L}}} ; \quad \underline{h}_{\text{eff}}\ket{\varPsi^{\text{R}}} = \eps\ket{\varPsi^{\text{R}}} ,
\ee
where the eigenvectors and eigenvalues correspond to the $4\times 4$ Nambu$\otimes$spin space. The solution for the TD1RDM expanded in the left eigenbasis takes the explicit form~\cite{Tuovinen2014-s}
\be
\langle {\varPsi}_j^{\text{L}} | \underline{\rho}(t) | {\varPsi}_k^{\text{L}} \rangle = \sum_\lambda \left\{{\varGamma}_{\lambda,jk} {\varLambda}_{\lambda,jk} + V_\lambda {\varGamma}_{\lambda,jk}\left[{\varPi}_{\lambda,jk}(t)+{\varPi}_{\lambda,kj}^*(t)\right] + V_\lambda^2{\varGamma}_{\lambda,jk}\ex^{-\im({\epsilon}_j-{\epsilon}_k^*)t}{\varOmega}_{\lambda,jk}\right\} , \label{eq:td1rdm}
\ee
where
\beq
{\varGamma}_{\lambda,jk} & = & \langle {\varPsi}_j^{\text{L}} | \underline{\varGamma}_\lambda | {\varPsi}_k^{\text{L}} \rangle , \\ \nonumber \\
{\varLambda}_{\lambda,jk} & = & \int\frac{\ud \omega}{2\pi} \frac{f(\w-\mu)}{(\w+V_\lambda-{\epsilon}_j)(\w+V_\lambda-{\epsilon}_k^*)} , \label{eq:lambda}\\ \nonumber \\
{\varPi}_{\lambda,jk}(t) & = & \int\frac{\ud \omega}{2\pi} \frac{f(\w-\mu)\ex^{\im(\w+V_\lambda-{\epsilon}_j)t}}{(\w-{\epsilon}_j)(\w+V_\lambda-{\epsilon}_j)(\w+V_\lambda-{\epsilon}_k^*)} , \label{eq:pi}\\ \nonumber \\
{\varOmega}_{\lambda,jk} & = & \int\frac{\ud \omega}{2\pi} \frac{f(\w-\mu)}{(\w-{\epsilon}_j)(\w+V_\lambda-{\epsilon}_j)(\w+V_\lambda-{\epsilon}_k^*)(\w-{\epsilon}_k^*)} . \nonumber \\ \label{eq:omega}
\eeq
Here $f(\w-\mu) = (\ex^{\beta(\w-\mu)} + 1)^{-1}$ is the Fermi function at inverse temperature $\beta$ and chemical potential $\mu$. Evaluating the TD1RDM in a physically relevant basis, e.g., the localized site basis of the central region $\{\ket{\varphi}\}$, is then readily done as a basis transformation from the left eigenbasis to the desired one
\be\label{eq:solution}
\mel{\varphi_m}{\underline{\rho}(t)}{\varphi_n} = \sum_{jk} \frac{\ev{\varphi_m}{{\varPsi}_j^{\text{R}}}}{\ev{{\varPsi}_j^{\text{L}}}{{\varPsi}_j^{\text{R}}}}
             \frac{\ev{{\varPsi}_k^{\text{R}}}{\varphi_n}}{\ev{{\varPsi}_k^{\text{R}}}{{\varPsi}_k^{\text{L}}}}
             \mel{{\varPsi}_j^{\text{L}}}{\underline{\rho}(t)}{{\varPsi}_k^{\text{L}}} ,
\ee
which follows from the biorthogonality of the left and right eigenvectors. The TD1RDM is then simply given by evaluating the terms in Eqs.~\eqref{eq:lambda}, \eqref{eq:pi} and~\eqref{eq:omega} for all indices $j,k$ and time parameter $t$, and then inserting into Eqs.~\eqref{eq:td1rdm} and~\eqref{eq:solution}.

The integrands in Eqs.~\eqref{eq:lambda}, \eqref{eq:pi} and~\eqref{eq:omega} have a fairly simple analytic structure: The ``$1/(\w-z)$'' type of terms have simple poles at $\w=z$ whereas the Fermi function has simple poles at the Matsubara frequencies given by $\w_n = (2n+1)\pi/(-\im\beta)$. Expressions similar to those in Eqs.~\eqref{eq:lambda}, \eqref{eq:pi}, \eqref{eq:omega} have been found, e.g., in Refs.~\cite{Jauho1994,Galperin2003} and integrated correspondingly using contour integration techniques. In Ref.~\cite{Tuovinen2014-s} the frequency integrals in Eqs.~\eqref{eq:lambda}, \eqref{eq:pi}, \eqref{eq:omega} were evaluated analytically in the zero-temperature limit to obtain a result for the TD1RDM in terms of logarithms and exponential integral functions. Here we evaluate these integrals analytically at arbitrary (inverse) temperature in the Fermi functions, and we will detail these steps next.

\section{Details of the Fermi integrals}
\begin{figure}[h!]
  \centering
  \includegraphics[width=0.25\textwidth]{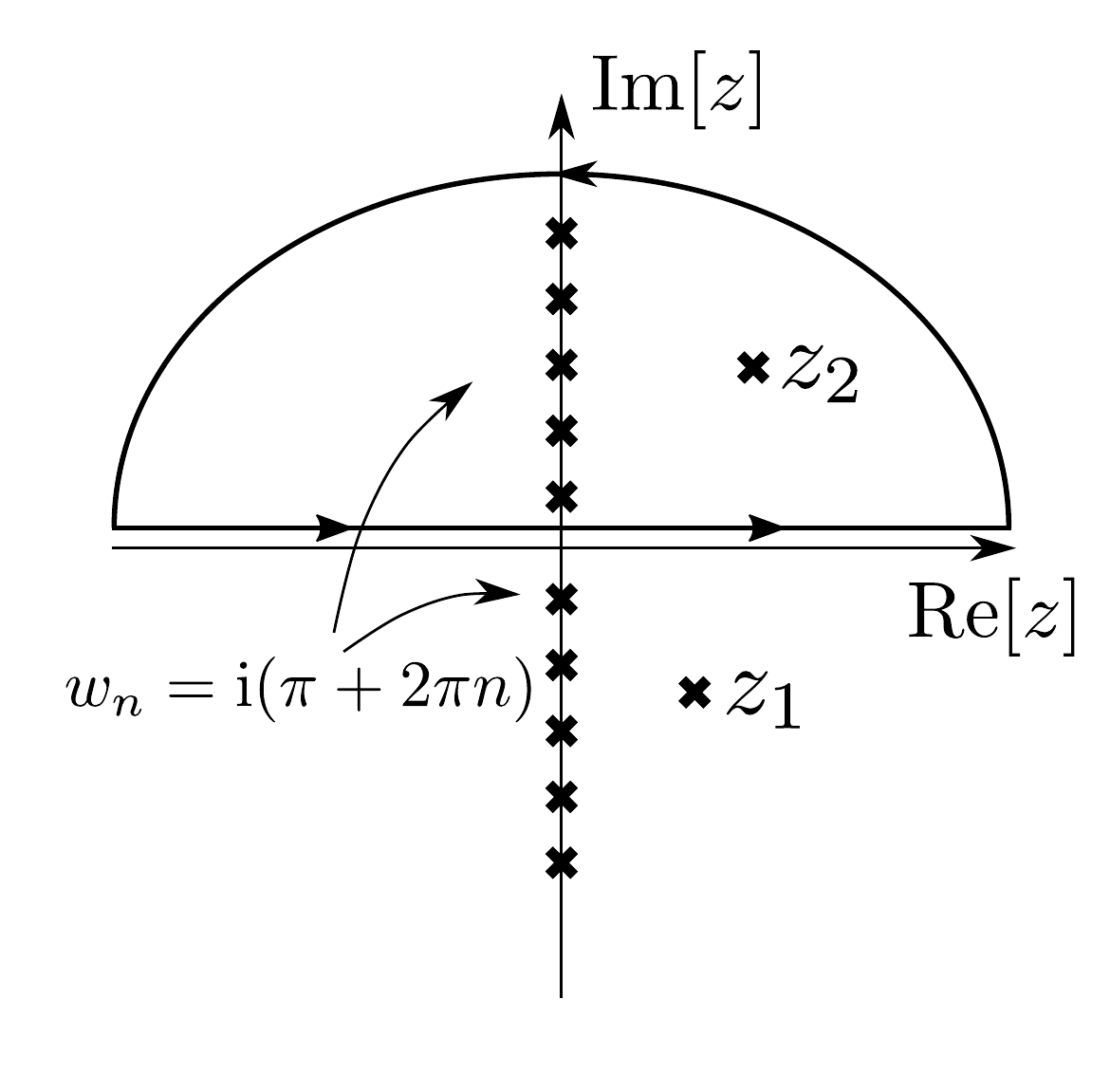}
  \caption{Poles in the complex $z$ plane for the integrand in Eq.~\eqref{eq:lambda_integral}. The locations of the poles are only for illustration.}
  \label{fig:poles}
\end{figure}

Making a change of variables $z=\beta(\w-\mu)$ in Eq.~\eqref{eq:lambda} gives 
\be\label{eq:lambda_integral}
{\varLambda}_{\lambda,jk} = \beta \int_{-\infty}^\infty \frac{\ud z}{2\pi} \frac{1}{(z-z_1)(z-z_2)(\ex^z + 1)} 
\ee
where we defined $z_1 = \beta({\epsilon}_j - \mu_\lambda)$ and
 $z_2 = \beta({\epsilon}_k^*-\mu_\lambda)$ with $\mu_\lambda = \mu + V_\lambda$. This integrand has simple poles at $z=z_1$, $z=z_2$ and $z = w_n = \im(\pi+2\pi n)$, see Fig.~\ref{fig:poles}. The spectrum of the complex eigenvalues of the nonhermitian matrix $\underline{h}_{\text{eff}}$ is such that the eigenvalues, ${\eps}_j$, lie in the lower-half plane (LHP) whereas the complex conjugated ones, ${\eps}_k^*$, lie in the upper-half plane (UHP).
For the ``$(z-z_n)^{-1}$'' contributions the residues are simply one and for the Fermi function we have $\text{Res}\left[(\ex^z+1)^{-1} \ , \ z=w_n\right] = -1 $.
Then, we can close the integral in Eq.~\eqref{eq:lambda_integral} in the UHP as shown in Fig.~\ref{fig:poles}, and using the residue theorem we get
\be\label{eq:lambda_integrated}
{\varLambda}_{\lambda,jk} = \im\beta\left[\frac{1}{z_2-z_1}\frac{1}{\ex^{z_2}+1} - \sum_{n=0}^\infty \frac{1}{(w_n-z_1)(w_n-z_2)}\right] .
\ee
The infinite sum can be written as
\beq
\sum_{n=0}^\infty \frac{1}{\left[\im(\pi + 2\pi n) - z_1\right]\left[\im(\pi + 2 \pi n) - z_2\right]} & = & \sum_{n=0}^\infty \frac{1}{2\pi \im \left(n + \frac{\im z_1 + \pi}{2\pi}\right) 2\pi \im \left(n + \frac{\im z_2 + \pi}{2\pi}\right)} = -\frac{1}{(2\pi)^2}\sum_{n=0}^\infty \frac{1}{(n+a)(n+b)} \nonumber \\
& = & -\frac{1}{(2\pi)^2}\frac{1}{b-a}\left[\psi(b) - \psi(a)\right]  ,
\eeq
where we defined $a=(\im z_1 + \pi)/2\pi$, $b = (\im z_2 + \pi)/2\pi$, and $\psi$ is the digamma function which is defined as the logarithmic derivative of the gamma function, $\psi(z) = \frac{\ud}{\ud z} \log \Gamma(z)$~\cite{digamma}. We can then insert the result of the sum back into Eq.~\eqref{eq:lambda_integrated} and couple the terms by simplifying
\beq\label{eq:lambda_finalresult-app}
{\varLambda}_{\lambda,jk} & = & \frac{\im}{{\epsilon}_k^* - {\epsilon}_j}\left\{\frac{1}{\ex^{\beta({\epsilon}_k^*-\mu_\lambda)}+1} + \frac{1}{2\pi\im}\left[\psi\left(\frac{1}{2}-\frac{\beta({\epsilon}_k^*-\mu_\lambda)}{2\pi\im}\right)-\psi\left(\frac{1}{2}-\frac{\beta({\epsilon}_j-\mu_\lambda)}{2\pi\im}\right)\right]\right\} 
\eeq
where we also inserted back the definitions of $z$'s. It is important to notice that we did not do anything but manipulations after using the residue theorem; the infinite sum was rewritten in terms of a special function $\psi$ which is broadly known in computational sciences and readily implemented for example in the GNU Scientific Library~\cite{gsl}. Eq.~\eqref{eq:lambda_finalresult-app} is our final result for ${\varLambda}_{\lambda,jk}$ for arbitrary values of $\beta$. We note in passing that it would give completely equivalent result if the integral was closed in the LHP.

Making the same change of variables in Eq.~\eqref{eq:pi} as in the previous case leads to
\be\label{eq:pi_integral}
{\varPi}_{\lambda,jk}(t) = \beta^2\int_{-\infty}^\infty \frac{\ud z}{2\pi}\frac{\ex^{\frac{\im}{\beta}(z-z_2)t}}{(z-z_1)(z-z_2)(z-z_3)(\ex^z + 1)}  ,
\ee
where we defined $z_1 = \beta({\epsilon}_j - \mu)$, $z_2 = \beta({\epsilon}_j - \mu_\lambda)$ and $z_3 = \beta({\epsilon}_k^* - \mu_\lambda)$. Also in this case we notice poles in the complex plane, similarly as in Fig.~\ref{fig:poles}. In this case, however, we may close the integral only in the UHP due to the exponential in the numerator, and we get according to the residue theorem
\be\label{eq:pi_integrated}
{\varPi}_{\lambda,jk}(t) = \im\beta^2\left[\frac{\ex^{\frac{\im}{\beta}(z_3-z_2)t}}{(z_3-z_1)(z_3-z_2)(\ex^{z_3}+1)} - \sum_{n=0}^\infty \frac{\ex^{\frac{\im}{\beta}(w_n-z_2)t}}{(w_n - z_1)(w_n - z_2)(w_n - z_3)}\right] . 
\ee
We may manipulate the infinite sum in Eq.~\eqref{eq:pi_integrated} as
\beq
\sum_{n=0}^\infty \frac{\ex^{\frac{\im}{\beta}(w_n-z_2)t}}{(w_n - z_1)(w_n - z_2)(w_n - z_3)} & = & \sum_{n=0}^\infty \frac{\ex^{\frac{\im}{\beta}2\pi\im\left(n+\frac{\im z_2+\pi}{2\pi}\right)t}}{2\pi\im\left(n + \frac{\im z_1+\pi}{2\pi}\right)2\pi\im\left(n + \frac{\im z_2+\pi}{2\pi}\right)2\pi\im\left(n + \frac{\im z_3+\pi}{2\pi}\right)} \nonumber \\
& = & \frac{\im}{(2\pi)^3}\sum_{n=0}^\infty \frac{\ex^{x(n+b)}}{(n+a)(n+b)(n+c)} ,
\eeq
where we defined $a=(\im z_1 + \pi)/2\pi$, $b=(\im z_2 + \pi)/2\pi$, $c=(\im z_3 + \pi)/2\pi$ and $x=-2\pi t/\beta$. In this case the infinite sum will give another type of special function, the hypergeometric function $\hypf$~\cite{hypf}:
\beq
& & \frac{\im}{(2\pi)^3}\sum_{n=0}^\infty \frac{\ex^{x(n+b)}}{(n+a)(n+b)(n+c)} \nonumber \\
& = & \frac{\im}{(2\pi)^3{(a-b)(a-c)(b-c)}}\left\{\ex^{bx}\left[ \frac{b-c}{a} \hypf(1,a,1+a,\ex^x) + \frac{c-a}{b} \hypf(1,b,1+b,\ex^x) + \frac{a-b}{c} \hypf(1,c,1+c,\ex^x)\right]\right\} . \nonumber \\
\eeq
The hypergeometric function together with the Pochhammer symbol are defined as~\cite{hypf,poch}
\be\label{eq:hypgeo}
\hypf(p,q,r,s) = \sum_{n=0}^\infty \frac{(p)_n (q)_n}{(r)_n}\frac{s^n}{n!} , \qquad (p)_n = \begin{cases}1 & n=1, \\ p(p+1)\cdots (p+n-1) & n>0.\end{cases} 
\ee
Inserting the definitions for $a$, $b$, $c$ and $x$ (and also the previously introduced variables $z$) leads to
\beq
& & \sum_{n=0}^\infty \frac{\ex^{\frac{\im}{\beta}(w_n-z_2)t}}{(w_n - z_1)(w_n - z_2)(w_n - z_3)} \nonumber \\
& = & \frac{-\im\ex^{-\pi t/\beta} \ex^{-\im({\epsilon}_j - \mu_\lambda)t}}{\beta^2({\eps}_k^* - {\eps}_j)({\eps}_k^* - {\eps}_j - V_\lambda)}\left\{ \mathfrak{{F}}({\eps}_k^*-\mu_\lambda,t,\beta) + \frac{{\eps}_k^*-{\eps}_j - V_\lambda}{V_\lambda} \mathfrak{{F}}({\eps}_j-\mu_\lambda,t,\beta) - \frac{{\eps}_k^* - {\eps}_j}{V_\lambda} \mathfrak{{F}}({\eps}_j-\mu,t,\beta)\right\} , 
\eeq
where we defined an auxiliary function
\be
\mathfrak{{F}}(z,t,\beta) \equiv \frac{1}{\im\beta z + \pi} \ \hypf\left(1, \frac{1}{2}+\frac{\im\beta z}{2\pi}, \frac{3}{2}+\frac{\im\beta z}{2\pi}, \ex^{-2\pi t/\beta}\right) . 
\ee
This calculation was only for the infinite sum in Eq.~\eqref{eq:pi_integrated}. Inserting the definitions of $z$'s into the first term gives
\be
\frac{\ex^{\frac{\im}{\beta}(z_3-z_2)t}}{(z_3-z_1)(z_3-z_2)(\ex^{z_3}+1)} = \frac{\ex^{-\im({\eps}_j - {\eps}_k^*)t}}{\beta^2({\eps}_k^* - {\eps}_j)({\eps}_k^* - {\eps}_j - V_\lambda)}\frac{1}{\ex^{\beta({\eps}_k^* - \mu_\lambda)}+1} .
\ee
Combining the terms finally gives
\beq\label{eq:pi_finalresult-app}
{\varPi}_{\lambda,jk}(t) & = & \frac{\im}{({\eps}_k^* - {\eps}_j)({\eps}_k^* - {\eps}_j - V_\lambda)}\left\{\frac{\ex^{-\im({\eps}_j - {\eps}_k^*)t}}{\ex^{\beta({\eps}_k^* - \mu_\lambda)}+1} + \im\ex^{-\pi t /\beta}\ex^{-\im({\eps}_j - \mu_\lambda)t}\times \right.\nonumber \\
& & \left.\left[\mathfrak{{F}}({\eps}_k^*-\mu_\lambda,t,\beta)  + \frac{{\eps}_k^*-{\eps}_j - V_\lambda}{V_\lambda} \mathfrak{{F}}({\eps}_j-\mu_\lambda,t,\beta) - \frac{{\eps}_k^* - {\eps}_j}{V_\lambda} \mathfrak{{F}}({\eps}_j-\mu,t,\beta)\right]\right\}
\eeq
for arbitrary values of $\beta$. Similarly here, after using the residue theorem, we only manipulated the expressions so that we could identify a known function $\hypf$. Conveniently, the hypergeometric function is also widely used in computational sciences, and both fast and accurate implementations of it are available~\cite{hypf-impl}.

In the third case, in Eq.~\eqref{eq:omega}, we do the same change of variables as before to get
\be\label{eq:omega_integral}
{\varOmega}_{\lambda,jk} = \beta^3\int_{-\infty}^\infty \frac{\ud z}{2\pi} \frac{1}{(z-z_1)(z-z_2)(z-z_3)(z-z_4)(\ex^z + 1)} ,
\ee
where we defined $z_1 = \beta({\eps}_j - \mu)$, $z_2 = \beta({\eps}_j - \mu_\lambda)$, $z_3 = \beta({\eps}_k^* - \mu_\lambda)$ and $z_4 = \beta({\eps}_k^* - \mu)$. The pole structure is again similar to the one shown in Fig.~\ref{fig:poles}, and we may close also this integral in the UHP. Again, according to the residue theorem we get as a result
\beq\label{eq:omega_integrated}
{\varOmega}_{\lambda,jk} & = & \im\beta^3\left\{\frac{1}{(z_3-z_1)(z_3-z_2)(z_3-z_4)(\ex^{z_3}+1)} + \frac{1}{(z_4-z_1)(z_4-z_2)(z_4-z_3)(\ex^{z_4}+1)} \right.\nonumber \\
& - & \left. \sum_{n=0}^\infty \frac{1}{(w_n-z_1)(w_n-z_2)(w_n-z_3)(w_n-z_4)}\right\} .
\eeq
The infinite sum may again be manipulated as
\beq
& & \sum_{n=0}^\infty\frac{1}{\left[\im(\pi + 2\pi n)-z_1\right]\left[\im(\pi + 2\pi n)-z_2\right]\left[\im(\pi + 2\pi n)-z_3\right]\left[\im(\pi + 2\pi n)-z_4\right]} \nonumber \\
& = & \sum_{n=0}^\infty\frac{1}{2\pi\im\left(n+\frac{\im z_1+\pi}{2\pi}\right)2\pi\im\left(n+\frac{\im z_2+\pi}{2\pi}\right)2\pi\im\left(n+\frac{\im z_3+\pi}{2\pi}\right)2\pi\im\left(n+\frac{\im z_4+\pi}{2\pi}\right)} \nonumber \\
& = & \frac{1}{(2\pi)^4}\sum_{n=0}^\infty\frac{1}{(n+a)(n+b)(n+c)(n+d)} ,
\eeq
where we defined $a=(\im z_1 + \pi)/2\pi$, $b=(\im z_2 + \pi)/2\pi$, $c=(\im z_3 + \pi)/2\pi$ and $d=(\im z_4 + \pi)/2\pi$. Also this sum has an expression in terms of the digamma function
\beq
\frac{1}{(2\pi)^4}\sum_{n=0}^\infty\frac{1}{(n+a)(n+b)(n+c)(n+d)} & = & \frac{1}{(2\pi)^4}\left[\frac{\psi(a)}{(a-b)(a-c)(a-d)} + \frac{\psi(b)}{(b-a)(b-c)(b-d)} \right.\nonumber \\
& + &\left.\frac{\psi(c)}{(c-a)(c-b)(c-d)} + \frac{\psi(d)}{(d-a)(d-b)(d-c)}\right] .
\eeq
Inserting the expressions for $a$, $b$, $c$ and $d$, and then further the expressions for $z_1$, $z_2$, $z_3$ and $z_4$ leads to
\beq
& & \sum_{n=0}^\infty \frac{1}{(w_n-z_1)(w_n-z_2)(w_n-z_3)(w_n-z_4)} \nonumber \\
& = & \frac{\im}{2\pi\beta^3}\left\{\frac{1}{({\eps}_k^*-{\eps}_j)({\eps}_k^* - {\eps}_j - V_\lambda)V_\lambda}\left[\psi\left(\frac{1}{2} + \frac{\im\beta({\eps}_j-\mu)}{2\pi}\right) - \psi\left(\frac{1}{2} + \frac{\im\beta({\eps}_k^*-\mu_\lambda)}{2\pi}\right)\right] \right.\nonumber \\
& - & \left. \frac{1}{({\eps}_k^*-{\eps}_j)({\eps}_k^* - {\eps}_j + V_\lambda)V_\lambda}\left[\psi\left(\frac{1}{2} + \frac{\im\beta({\eps}_j-\mu_\lambda)}{2\pi}\right) - \psi\left(\frac{1}{2} + \frac{\im\beta({\eps}_k^*-\mu)}{2\pi}\right)\right]\right\} .
\eeq
Combining the terms in Eq.~\eqref{eq:omega_integrated} gives as the final result
\beq\label{eq:omega_finalresult-app}
& & {\varOmega}_{\lambda,jk} \nonumber \\
& = & \frac{\frac{\im}{\ex^{\beta({\eps}_k^*-\mu)}+1}-\frac{1}{2\pi}\left[\psi\left(\frac{1}{2} - \frac{\beta({\eps}_j-\mu_\lambda)}{2\pi\im}\right) - \psi\left(\frac{1}{2} - \frac{\beta({\eps}_k^*-\mu)}{2\pi\im}\right)\right]}{({\eps}_k^*-{\eps}_j)({\eps}_k^* - {\eps}_j + V_\lambda)V_\lambda} - \frac{\frac{\im}{\ex^{\beta({\eps}_k^*-\mu_\lambda)}+1}-\frac{1}{2\pi}\left[\psi\left(\frac{1}{2} - \frac{\beta({\eps}_j-\mu)}{2\pi\im}\right) - \psi\left(\frac{1}{2} - \frac{\beta({\eps}_k^*-\mu_\lambda)}{2\pi\im}\right)\right]}{({\eps}_k^*-{\eps}_j)({\eps}_k^* - {\eps}_j - V_\lambda)V_\lambda} \nonumber \\
\eeq
for arbitrary values of $\beta$.

Finally, inserting Eqs.~\eqref{eq:lambda_finalresult-app}, \eqref{eq:pi_finalresult-app} and~\eqref{eq:omega_finalresult-app} into Eq.~\eqref{eq:td1rdm} gives then the TD1RDM at arbitrary temperature. When the asymptotic behaviour of the digamma and hypergeometric function is studied, the results in Eqs.~\eqref{eq:lambda_finalresult-app}, \eqref{eq:pi_finalresult-app} and~\eqref{eq:omega_finalresult-app} can be shown to reduce to those in Ref.~\cite{Tuovinen2014-s} at the zero-temperature limit ($\beta\to\infty$)~\cite{Tuovinen2016thesis}. We also note that congruent results involving equivalent special functions have recently been reported in Refs.~\cite{RidleyPNGF6,Ridley2016,PhysRevB.95.165440}.

\section{Inclusion of sudden electromagnetic fields in the central region}
It is also possible to include a sudden switch-on of an electromagnetic field in the Hamiltonian of the central region. For example, this includes the possibility for a static potential profile (e.g.\ a gate voltage) $u_{mn}$, between basis states $m,n$ of the central region, to be added to the ``on-site'' contribution $\underline{a}$ [Eq.~\eqref{eq:onsite}]. Also, for the ``nearest-neighbor'' contribution $\underline{b}$ [Eq.~\eqref{eq:nn}], it is possible to consider a Peierls phase $\gamma_{mn} = -\gamma_{nm}$ accounting for a magnetic field (normalized to the flux quantum $\phi_0 = h/2e$) when traversed along a closed loop of states $m,n$. For a general description, we simply consider a perturbed Hamiltonian $\widetilde{\underline{h}}_{CC}$ out of equilibrium (signified by a tilde), and use the unperturbed Hamiltonian $\underline{h}_{CC}$ in equilibrium. Then, a formula for the TD1RDM similar to Eq.~\eqref{eq:td1rdm} can be derived as~\cite{Tuovinen2014-s}
\be
\mel{\widetilde{{\varPsi}}_j^{\text{L}}}{\widetilde{\underline{\rho}}(t)}{\widetilde{{\varPsi}}_k^{\text{L}}} = \sum_\lambda\left[\widetilde{{\varGamma}}_{\lambda,jk}\widetilde{{\varLambda}}_{\lambda,jk} + \widetilde{{\varPi}}_{\lambda,jk}(t) + \widetilde{{\varPi}}_{\lambda,kj}^*(t) + \widetilde{{\varOmega}}_{\lambda,jk}(t)\right] , \label{eq:td1rdm-pert} 
\ee
where the introduced terms $\widetilde{\varGamma}$, $\widetilde {\varPi}$ and $\widetilde{\varOmega}$ take a slightly more intricate form compared to those in Eq.~\eqref{eq:td1rdm} as the eigenbases of the unperturbed and perturbed Hamiltonians, in general, do not need to be the same. Therefore, we need to take the corresponding overlaps into account
\beq
\widetilde{{\varGamma}}_{\lambda,jk} & = & \mel{\widetilde{{\varPsi}}_j^{\text{L}}}{\underline{\varGamma}_{\lambda}}{\widetilde{{\varPsi}}_k^{\text{L}}} , \\ \nonumber \\
\widetilde{{\varLambda}}_{\lambda,jk} & = & \int\frac{\ud \omega}{2\pi}\frac{f(\omega - \mu)}{(\w+V_{\lambda}-\widetilde{{\eps}}_j)(\w+V_{\lambda}-\widetilde{{\eps}}_k^*)} , \label{eq:lambda-pert}\\ \nonumber \\
\widetilde{{\varPi}}_{\lambda,jk}(t) & = & \sum\limits_{mn}\frac{\ev{\widetilde{{\varPsi}}_j^{\text{L}}}{{\varPsi}_m^{\text{R}}}\mel{{\varPsi}_m^{\text{L}}}{\widetilde{\underline{V}}_{\lambda}}{\widetilde{{\varPsi}}_n^{\text{R}}}\widetilde{{\varGamma}}_{\lambda,nk}}{\ev{{\varPsi}_m^{\text{L}}}{{\varPsi}_m^{\text{R}}}\ev{\widetilde{{\varPsi}}_n^{\text{L}}}{\widetilde{{\varPsi}}_n^{\text{R}}}} \int\frac{\ud \omega}{2\pi}\frac{f(\omega - \mu)\ex^{\im(\w+V_{\lambda}-\widetilde{{\eps}}_j)t}}{(\w-{\eps}_m)(\w+V_{\lambda}-\widetilde{{\eps}}_n)(\w+V_{\lambda}-\widetilde{{\eps}}_k^*)} , \label{eq:pi-pert} \\ \nonumber \\
\widetilde{{\varOmega}}_{\lambda,jk}(t) & = & \sum\limits_{mnpq}\frac{\ev{\widetilde{{\varPsi}}_j^{\text{L}}}{{\varPsi}_m^{\text{R}}}\mel{{\varPsi}_m^{\text{L}}}{\widetilde{\underline{V}}_{\lambda}}{\widetilde{{\varPsi}}_n^{\text{R}}}\widetilde{{\varGamma}}_{\lambda,np}\mel{\widetilde{{\varPsi}}_p^{\text{R}}}{\widetilde{\underline{V}}_{\lambda}^\dagger}{{\varPsi}_q^{\text{L}}}\ev{{\varPsi}_q^{\text{R}}}{\widetilde{{\varPsi}}_k^{\text{L}}}}{\ev{{\varPsi}_m^{\text{L}}}{{\varPsi}_m^{\text{R}}}\ev{\widetilde{{\varPsi}}_n^{\text{L}}}{\widetilde{{\varPsi}}_n^{\text{R}}}\ev{\widetilde{{\varPsi}}_p^{\text{R}}}{\widetilde{{\varPsi}}_p^{\text{L}}}\ev{{\varPsi}_q^{\text{R}}}{{\varPsi}_q^{\text{L}}}} \nonumber \\
& \times & \ex^{-\im(\widetilde{{\eps}}_j-\widetilde{{\eps}}_k^*)t}\int\frac{\ud \omega}{2\pi}\frac{f(\omega - \mu)}{(\w-{\eps}_m)(\w+V_{\lambda}-\widetilde{{\eps}}_n)(\w+V_{\lambda}-\widetilde{{\eps}}_p^*)(\w-{\eps}_q^*)} ,  \label{eq:omega-pert}
\eeq
where the tildes signify that the corresponding quantities are calculated from the perturbed Hamiltonian $\widetilde{\underline{h}}_{CC}$, and we explicitly defined a ``bias-voltage matrix'' $\widetilde{\underline{V}}_\lambda \equiv V_{\lambda}\underline{\unitmat} - (\widetilde{\underline{h}}_{CC} - \underline{h}_{CC})$. The eigenvalues $\{{\eps},\widetilde{{\eps}}\}$ and eigenvectors $\{{\varPsi}^{\text{L/R}},\widetilde{{\varPsi}}^{\text{L/R}}\}$ refer to the complex eigenvalues and to the left/right eigenvectors of $\underline{h}_{\mathrm{eff}}$ and $\widetilde{\underline{h}}_{\mathrm{eff}}=\widetilde{\underline{h}}_{CC}-\im\underline{\varGamma}/2$, respectively. In the limit $\widetilde{\underline{h}}_{CC} \to \underline{h}_{CC}$ the result in Eq.~\eqref{eq:td1rdm-pert} can be checked to reduce to Eq.~\eqref{eq:td1rdm}~\cite{Tuovinen2016thesis}.

Similarly, for the TD1RDM with sudden electromagnetic fields in the central region in Eq.~\eqref{eq:td1rdm-pert}, we can take the integrals in Eqs.~\eqref{eq:lambda-pert}, \eqref{eq:pi-pert} and~\eqref{eq:omega-pert} and evaluate them in the same manner. This time the pole structure is only a little more intricate due to different eigenvalues for the unperturbed and perturbed Hamiltonians but it can be handled exactly in the same way as above. For perturbed central regions at arbitrary $\beta$ the explicit results are
\beq
\widetilde{{\varLambda}}_{\lambda,jk} & = & \frac{\im}{\widetilde{{\epsilon}}_k^* - \widetilde{{\epsilon}}_j}\left\{\frac{1}{\ex^{\beta(\widetilde{{\epsilon}}_k^*-\mu_\lambda)}+1} + \frac{1}{2\pi\im}\left[\psi\left(\frac{1}{2}-\frac{\beta(\widetilde{{\epsilon}}_k^*-\mu_\lambda)}{2\pi\im}\right)-\psi\left(\frac{1}{2}-\frac{\beta(\widetilde{{\epsilon}}_j-\mu_\lambda)}{2\pi\im}\right)\right]\right\} , \label{eq:lambda_finalresult-pert} \\ \nonumber \\
\widetilde{{\varPi}}_{\lambda,jk}(t) & = & \sum\limits_{mn}\frac{\ev{\widetilde{{\varPsi}}_j^{\text{L}}}{{\varPsi}_m^{\text{R}}}\mel{{\varPsi}_m^{\text{L}}}{\widetilde{\underline{V}}_{\lambda}}{\widetilde{{\varPsi}}_n^{\text{R}}}\widetilde{{\varGamma}}_{\lambda,nk}}{\ev{{\varPsi}_m^{\text{L}}}{{\varPsi}_m^{\text{R}}}\ev{\widetilde{{\varPsi}}_n^{\text{L}}}{\widetilde{{\varPsi}}_n^{\text{R}}}} \times \nonumber \\
& & \frac{\im}{(\widetilde{{\eps}}_k^* - \widetilde{{\eps}}_n)(\widetilde{{\eps}}_k^* - {\eps}_m - V_\lambda)}\left\{\frac{\ex^{-\im(\widetilde{{\eps}}_j - \widetilde{{\eps}}_k^*)t}}{\ex^{\beta(\widetilde{{\eps}}_k^* - \mu_\lambda)}+1} + \im\ex^{-\pi t /\beta}\ex^{-\im(\widetilde{{\eps}}_j - \mu_\lambda)t}\times\right.\nonumber \\
&  & \left. \left[\mathfrak{{F}}(\widetilde{{\eps}}_k^*-\mu_\lambda,t,\beta) - \frac{\widetilde{{\eps}}_k^*-{\eps}_m - V_\lambda}{\widetilde{{\eps}}_n -{\eps}_m - V_\lambda} \mathfrak{{F}}(\widetilde{{\eps}}_n-\mu_\lambda,t,\beta) + \frac{\widetilde{{\eps}}_k^* - \widetilde{{\eps}}_n}{\widetilde{{\eps}}_n - {\eps}_m - V_\lambda} \mathfrak{{F}}({\eps}_m-\mu,t,\beta)\right]\right\} , \label{eq:pi_finalresult-pert} \\ \nonumber \\ 
\widetilde{{\varOmega}}_{\lambda,jk} & = & \sum\limits_{mnpq}\frac{\ev{\widetilde{{\varPsi}}_j^{\text{L}}}{{\varPsi}_m^{\text{R}}}\mel{{\varPsi}_m^{\text{L}}}{\widetilde{\underline{V}}_{\lambda}}{\widetilde{{\varPsi}}_n^{\text{R}}}\widetilde{{\varGamma}}_{\lambda,np}\mel{\widetilde{{\varPsi}}_p^{\text{R}}}{\widetilde{\underline{V}}_{\lambda}^\dagger}{{\varPsi}_q^{\text{L}}}\ev{{\varPsi}_q^{\text{R}}}{\widetilde{{\varPsi}}_k^{\text{L}}}}{\ev{{\varPsi}_m^{\text{L}}}{{\varPsi}_m^{\text{R}}}\ev{\widetilde{{\varPsi}}_n^{\text{L}}}{\widetilde{{\varPsi}}_n^{\text{R}}}\ev{\widetilde{{\varPsi}}_p^{\text{R}}}{\widetilde{{\varPsi}}_p^{\text{L}}}\ev{{\varPsi}_q^{\text{R}}}{{\varPsi}_q^{\text{L}}}} \ \ex^{-\im(\widetilde{{\eps}}_j-\widetilde{{\eps}}_k^*)t} \times \nonumber \\
& & \left\{\frac{1}{({\eps}_m - \widetilde{{\eps}}_n + V_\lambda)({\eps}_m - \widetilde{{\eps}}_p^* + V_\lambda)({\eps}_m - {\eps}_q^*)}\frac{1}{2\pi}\psi\left(\frac{1}{2} - \frac{\beta({\eps}_m-\mu)}{2\pi\im}\right) \right.\nonumber \\
& + & \left.\frac{1}{(\widetilde{{\eps}}_n - {\eps}_m - V_\lambda)(\widetilde{{\eps}}_n - \widetilde{{\eps}}_p^*)(\widetilde{{\eps}}_n - {\eps}_q^* - V_\lambda)}\frac{1}{2\pi}\psi\left(\frac{1}{2} - \frac{\beta(\widetilde{{\eps}}_n-\mu_\lambda)}{2\pi\im}\right) \right.\nonumber \\ 
& + & \left.\frac{1}{({\eps}_q^* - {\eps}_m)({\eps}_q^* - \widetilde{{\eps}}_n + V_\lambda)({\eps}_q^* - \widetilde{{\eps}}_p^* + V_\lambda)}\left[\frac{\im}{\ex^{\beta({\eps}_q^* - \mu)}+1} + \frac{1}{2\pi}\psi\left(\frac{1}{2} - \frac{\beta({\eps}_q^*-\mu)}{2\pi\im}\right)\right] \right.\nonumber \\
& + & \left.\frac{1}{(\widetilde{{\eps}}_p^* - {\eps}_m - V_\lambda)(\widetilde{{\eps}}_p^* - \widetilde{{\eps}}_n)(\widetilde{{\eps}}_p^* - {\eps}_q^* - V_\lambda)}\left[\frac{\im}{\ex^{\beta(\widetilde{{\eps}}_p^* - \mu_\lambda)}+1} + \frac{1}{2\pi}\psi\left(\frac{1}{2} - \frac{\beta(\widetilde{{\eps}}_p^*-\mu_\lambda)}{2\pi\im}\right)\right]\right\} . \label{eq:omega_finalresult-pert}
\eeq
Again, inserting Eqs.~\eqref{eq:lambda_finalresult-pert}, \eqref{eq:pi_finalresult-pert} and~\eqref{eq:omega_finalresult-pert} into Eq.~\eqref{eq:td1rdm-pert} gives then the TD1RDM for a perturbed central region at arbitrary temperature. Also here, the zero-temperature limit ($\beta\to\infty$) presented in Ref.~\cite{Tuovinen2014-s}, is recovered by the asymptotics of the digamma and hypergeometric functions in Eqs.~\eqref{eq:lambda_finalresult-pert}, \eqref{eq:pi_finalresult-pert} and~\eqref{eq:omega_finalresult-pert}~\cite{Tuovinen2016thesis}. By careful inspection of Eqs.~\eqref{eq:lambda_finalresult-pert}, \eqref{eq:pi_finalresult-pert} and~\eqref{eq:omega_finalresult-pert} in the limit of unperturbed central region ($\widetilde{{\varPsi}}\to{\varPsi}$ and $\widetilde{{\eps}}\to{\eps}$) it can be verified that they reduce to Eqs.~\eqref{eq:lambda_finalresult-app}, \eqref{eq:pi_finalresult-app} and~\eqref{eq:omega_finalresult-app}~\cite{Tuovinen2016thesis}.

\section{Derivation of the bond current}
We define the bond current flowing between site $j$ and $j+1$ in the nanowire (central device) by the rate of change of the number of particles in the region comprising the left electrode and the first $j$ sites in the nanowire:
\be
\hat{N}_j = \sum_{ks} \hat{c}_{kLs}^\dagger \hat{c}_{kLs} + \sum_{m=1}^j \sum_s \hat{c}_{ms}^\dagger \hat{c}_{ms},
\ee
where $k$ and $m$ respectively label the basis elements in the left electrode and the sites in the nanowire, and $s$ is a spin index. The bond current between sites $j$ and $j+1$ is then defined by
\be\label{eq:current}
I_{j,j+1} \equiv \frac{\ud}{\ud t} \langle\hat{N}_j\rangle .
\ee
The temporal change in the number of particles can be derived from
\be\label{eq:commutator}
\frac{\ud}{\ud t} \langle\hat{N}_j\rangle = -\im \left\langle\left[\hat{N}_j , \hat{H}_{\text{total}}\right]\right\rangle,
\ee
where we now separate the `normal' and `superconducting' contributions as $\hat{H}_{\text{total}} = \hat{H}_{\text{normal}} + \hat{H}_\varDelta$ with
\beq
\hat{H}_{\text{normal}} & = & \hat{H}_L + \hat{H}_R + \hat{H}_c + \sum_{n=1}^{N_w} \sum_{ss'}\left(\epsilon^{ss'}\hat{c}_{ns}^\dagger \hat{c}_{ns'} + \hc\right) + \sum_{n=1}^{N_w-1}\sum_{ss'}\left(J^{ss'}\hat{c}_{ns}^\dagger \hat{c}_{(n+1)s'}+\hc\right), \\
\hat{H}_\varDelta & = & \varDelta \sum_{n=1}^{N_w} \hat{c}_{n\uparrow}\hat{c}_{n\downarrow} + \hc ,
\eeq
where we separated the ``on-site'' and ``nearest-neighbor'' contributions in the spin-dependent matrix elements of $\epsilon$ and $J$. Also, the nanowire is coupled to the electrodes only via the terminal sites ($1$ and $N_w$), so the coupling Hamiltonian takes the form
\be
\hat{H}_c = \sum_{kss'} \left[\left(T_{kL1}^{ss'} \hat{c}_{kLs}^\dagger \hat{c}_{1s'} + \hc \right) + \left(T_{kRN_w}^{ss'} \hat{c}_{kRs}^\dagger \hat{c}_{N_w s'} + \hc \right) \right].
\ee

With the normal part of the Hamiltonian, the commutator in Eq.~\eqref{eq:commutator} is nonzero only for the following terms
\beq
\left[\hat{N}_j , \hat{H}_{\text{normal}}\right] & = & \left[\sum_{ks} \hat{c}_{kLs}^\dagger \hat{c}_{kLs},\hat{H}_c\right] + \left[\sum_{m=1}^j \sum_s \hat{c}_{ms}^\dagger \hat{c}_{ms},\hat{H}_c\right] \nonumber \\
& + & \left[\sum_{m=1}^j \sum_s \hat{c}_{ms}^\dagger \hat{c}_{ms} , \sum_{n=1}^{N_w-1}\sum_{s's''}\left(J^{s's''}\hat{c}_{ns'}^\dagger \hat{c}_{(n+1)s''}+ J^{s''s'} \hat{c}_{(n+1)s''}^\dagger \hat{c}_{ns'} \right)\right] .
\eeq
As the coupling Hamiltonian $\hat{H}_c$ has one creation (annihilation) operator in the nanowire and one annihilation (creation) operator in the electrode, so in principle the first two terms above can give a nonzero commutator, but it turns out they cancel each other out. We are then left with the term on the second line which can be simplified to give
\be\label{eq:terma}
\left[\hat{N}_j , \hat{H}_{\text{normal}}\right] = \sum_{ss'} J^{ss'} \left(\hat{c}_{js}^\dagger \hat{c}_{(j+1)s'} - \hat{c}_{(j+1)s}^\dagger \hat{c}_{js'}\right).
\ee
The remaining calculation is the commutator with the `superconducting' part where the nonzero contribution comes from
\be\label{eq:termb}
\left[\hat{N}_j , \hat{H}_\varDelta\right] = \left[\sum_{m=1}^j \sum_s \hat{c}_{ms}^\dagger \hat{c}_{ms},  \varDelta \sum_{n=1}^{N_w} \hat{c}_{n\uparrow}\hat{c}_{n\downarrow} + \hc\right] = \sum_{m=1}^j\left( \varDelta \hat{c}_{m\downarrow} \hat{c}_{m\uparrow} - \varDelta \hat{c}_{m\uparrow} \hat{c}_{m\downarrow} - \varDelta^* \hat{c}_{m\uparrow}^\dagger\hat{c}_{m\downarrow}^\dagger + \varDelta^* \hat{c}_{m\downarrow}^\dagger\hat{c}_{m\uparrow}^\dagger \right).
\ee
We may then insert Eqs.~\eqref{eq:terma} and~\eqref{eq:termb} into Eq.~\eqref{eq:commutator} and further into Eq.~\eqref{eq:current}, and use the model parameters for the nanowire (we have assumed a real pairing field) to obtain:
\beq
I_{j,j+1} & = & -\im \left\langle -\frac{J}{2} \left[\hat{c}_{j\uparrow}^\dagger \hat{c}_{(j+1)\uparrow} + \hat{c}_{j\downarrow}^\dagger \hat{c}_{(j+1)\downarrow} - \left( \hat{c}_{(j+1)\uparrow}^\dagger \hat{c}_{j\uparrow} + \hat{c}_{(j+1)\downarrow}^\dagger \hat{c}_{j\downarrow} \right) \right] \right.\nonumber \\
& - & \left.\frac{\alpha}{2} \left[\hat{c}_{j\uparrow}^\dagger \hat{c}_{(j+1)\downarrow} - \hat{c}_{j\downarrow}^\dagger \hat{c}_{(j+1)\uparrow} - \left( \hat{c}_{(j+1)\downarrow}^\dagger \hat{c}_{j\uparrow} - \hat{c}_{(j+1)\uparrow}^\dagger \hat{c}_{j\downarrow} \right) \right] \right. \nonumber \\
& + & \left. \sum_{m=1}^j \varDelta \left( \hat{c}_{m\downarrow} \hat{c}_{m\uparrow} - \hat{c}_{m\uparrow} \hat{c}_{m\downarrow} - \hat{c}_{m\uparrow}^\dagger\hat{c}_{m\downarrow}^\dagger + \hat{c}_{m\downarrow}^\dagger\hat{c}_{m\uparrow}^\dagger \right) \right\rangle .
\eeq
This may further be simplified as in Eq.~\eqref{eq:bondcurrent} in the main text.

%

\end{document}